\documentclass[utf8,11pt]{frontiersSCNS}
\usepackage{url,hyperref,microtype,subcaption}
\usepackage[onehalfspacing]{setspace}

\def\keyFont{\fontsize{8}{11}\helveticabold }
\def\firstAuthorLast{Sample {et~al.}} 
\def\Authors{Paulina Lira\,$^{1}$, Ismael Botti\,$^{2}$, Shai Kaspi\,$^{3}$ and Hagai Netzer\,$^{3,4}$}

\def\Address{$^{1,*}$Departamento de Astronom\'ia, Universidad de Chile, Casilla 36D, Santiago, Chile\\
  $^{2}$School of Department of Physics and Astronomy, University of Nottingham, University Park, Nottingham NG7 2RD, UK\\
  $^{3}$School of Physics and Astronomy, Tel Aviv University, Tel Aviv 69978, Israel\\
  $^{4}$Wise Observatory, School of Physics and Astronomy, Tel Aviv University, Tel Aviv 69978, Israel}

\def\corrAuthor{Paulina Lira}
\def\corrEmail{plira@das.uchile.cl}

\begin{document}
\onecolumn
\firstpage{1}

\title[Reverberation Mapping of Quasars]{Reverberation Mapping of High-$z$, High-luminosity Quasars}

\author[\firstAuthorLast ]{\Authors} 
\address{} 
\correspondance{} 
\extraAuth{}

\maketitle

\begin{abstract}

We present Reverberation Mapping results after monitoring a sample of
17 high-$z$, high-luminosity quasars for more than 10 years using
photometric and spectroscopic capabilities. Continuum and line
emission flux variability is observed in all quasars. Using
cross-correlation analysis we successfully determine lags between the
variations in the continuum and broad emission lines for several
sources.  Here we present a highlight of our results and the
determined radius--luminosity relations for Ly$\alpha$ and CIV.

\tiny
 \keyFont{\section{Keywords:} Quasars, Black Holes, Reverberation Mapping, Broad Line Region, AGN}
\end{abstract}

\section{Introduction}

Reverberation Mapping (RM) has been an extremely successful technique
used to study the innermost regions of Active Galactic Nuclei
(AGN). The determination of lags between variations in the continuum
emission coming from the accretion disk near the central Black Hole
(BH), and the response from the emission lines produced in the Broad
Line Region (BLR) have shown that the BLR is an extended, virialized
and ionizion-stratified structure. Furthermore, the determination of
the radius--luminosity relation between the BLR radius at which
H$\beta$ is produced and the luminosity of the central source has
open, through cross-calibration to other BLR lines, the possibility to
measure BH masses in hundred of thousands of sources. The
cross-calibration is, however, subject to many uncertainties due to
the extrapolations necessary to apply the radius--luminosity relation
to sources of very different luminosities to those actually probed
with RM experiments, and to the use of emission lines produced by
regions of the BLR that can be far from that producing H$\beta$. This
is the motivation to conduct RM campaigns in high-$z$, high-luminosity
quasars for those emission lines available in the observed optical
domain.

\section{Observational campaign and resulting light curves}

Since 2005 we undertook a long observational campaign to monitor a
sample of southern, high-redshift ($z \sim 2.5 -3$, with one source at
$z = 1.8$), high-luminosity ($M_B \sim -29$) quasars. 50 targets were
originally selected from the SDSS \citep{2005AJ....130..367S} and
Cal\'an-Tololo samples \citep{1996RMxAA..32...35M}. We started with a
purely R-band imaging (corresponding to rest frame wavelengths $\sim
1700-1800$\AA, depending on the exact redshift of the source) and two
years later triggered the first spectroscopic observations of those
quasars with the largest photometric variations. Over the years the
campaign was narrowed down to a final sample of 17 quasars which have
good quality R-band and emission line light curves.

Line fluxes were measured using spectral windows tailored to each line
and each quasar. We avoided regions where the lines were contaminated
by variable absorbing features, but did not attempt to correct for the
contribution of other (weaker) emission lines, either narrow or
broad. For further details see Lira et al.~(2018, submitted).

Over the campaign we found that most quasars showed a substantial
degree of variability in the continuum and line emission line fluxes,
with typical normalized variability amplitudes $f_{\rm var}$
\citep{1997ApJS..110....9R} of $\sim 10$\%. Most quasars also showed
the expected behavior, where the emission lines followed the trends
seen in the continuum as traced by the R-band light curves. Many
sources, unfortunately, did not shown enough structure in their light
curves (which in the rest frame only map the continuum and line flux
variations during $\sim 3$ years), to allow for statically significant
lag determinations.

Two sources showed unexpected Ly$\alpha$ light curves, where the line
fluxes depart from the behavior shown by the continuum and the
remaining emission lines. One example (J224743) is shown in Figure 1,
where we also include a source that presents the expected line
response to the continuum variations (CT650).

\begin{figure}[h!]
\begin{center}
\includegraphics[scale=0.5]{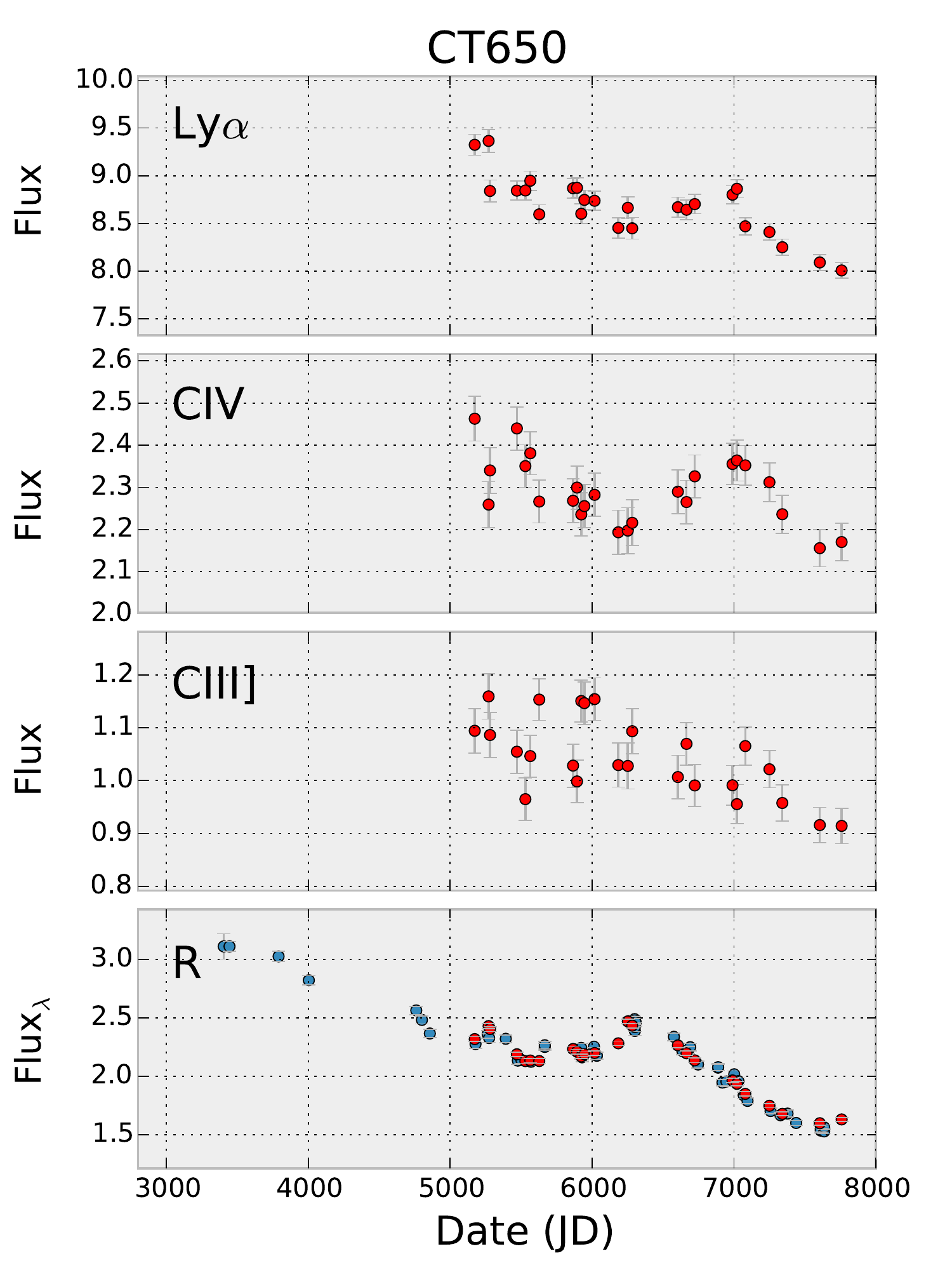}%
\includegraphics[scale=0.5]{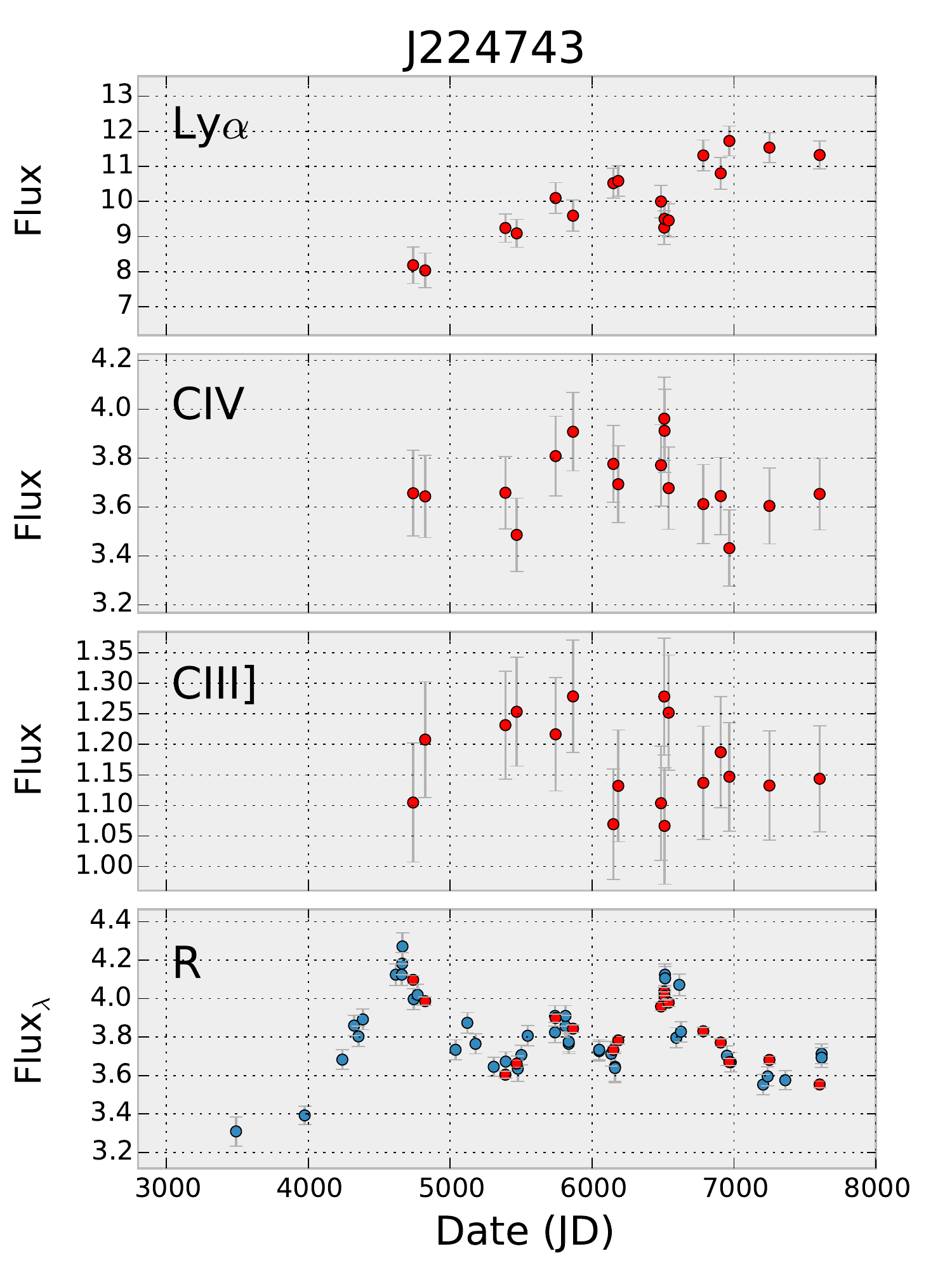}
\end{center}
\caption{R-band and emission line light curves for a well behaved
  source, CT650 (left), and an anomalous quasar, J224743 (right). The
  time axis is expressed in Julian Dates - 2450000 days. Blue points
  correspond to photometric observations while red points were
  obtained from the spectroscopic data. The continuum flux
  measurements shown in the bottom panel, either photometric or
  spectroscopic, correspond to the observer-frame R-band, and
  therefore the corresponding rest-frame spectral coverage changes
  from quasar to quasar. Taken from Lira et al.~(submitted),
  \copyright AAS.}
\end{figure}

\section{Time series analysis}

Cross-correlation analysis was conducted using the ICCF and ZCCF
methods \citep{1986ApJ...305..175G,1997ASSL..218..163A}. Errors were
determined using Monte Carlo simulations were the light curve fluxes
were randomized and bootstrapped to contain about 70\%\ of the
original data points \citep{2004ApJ...613..682P}. We determined
statistically significant lags with respect to that of the R-band
continuum for 3 Ly$\alpha$, 5 CIV, 1 SiIV, 1 CIII], and 1 MgII
  emission line light curves.

In Figure 2 we reproduce the light curves for two of our sources,
CT286 and J221516. The emission line light curves have been shifted
according to the calculated lags, while all curves have been taken to
a mean of zero and a standard deviation of one. Figure 2 illustrates
that emission line light curves can closely match the observed UV
continuum, like in the case of J221516, or can show rather different
trends, as seen in CT286, where the line light curves correspond to a
heavily smoothed version of the continuum light curve.

Two reasons can be invoked to explain such differences: 1) the
response of the BLR to continuum variations differs from object to
object, and it can be nonlinear and show variations with time; 2) the
observed UV continuum might not be a good representation of the
ionizing continuum, which is actually responsible of driving the
emission line changes.

The recent monitoring of NGC5548 by the STORM consortium displays
several of these traits during the 170 days of monitoring
\citep{2016ApJ...824...11G}. Emission line light curves follow closely
that of the continuum during the first 1/3 of the campaign, to then
disengage from it showing a decorrelated behavior for $\sim 60-70$
days, to finally going back to the original state. Besides, while some
emission lines show a smoother light curve than that of the observed
continuum during the last segment of the monitoring, SiIV stands out
for showing larger amplitude in its peaks and troughs than that of the
continuum.

\begin{figure}[h!]
\begin{center}
\includegraphics[scale=0.4]{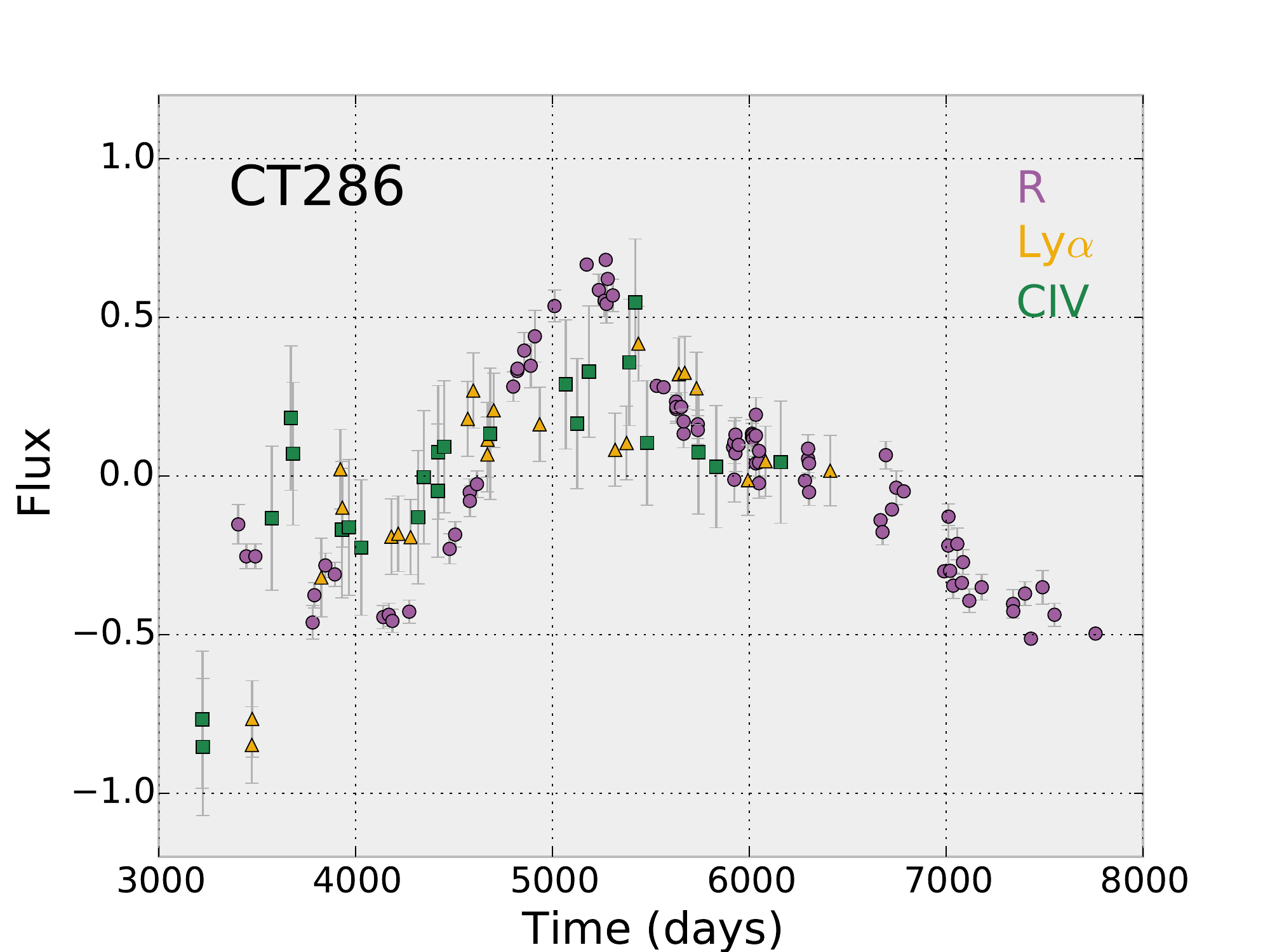}%
\includegraphics[scale=0.4]{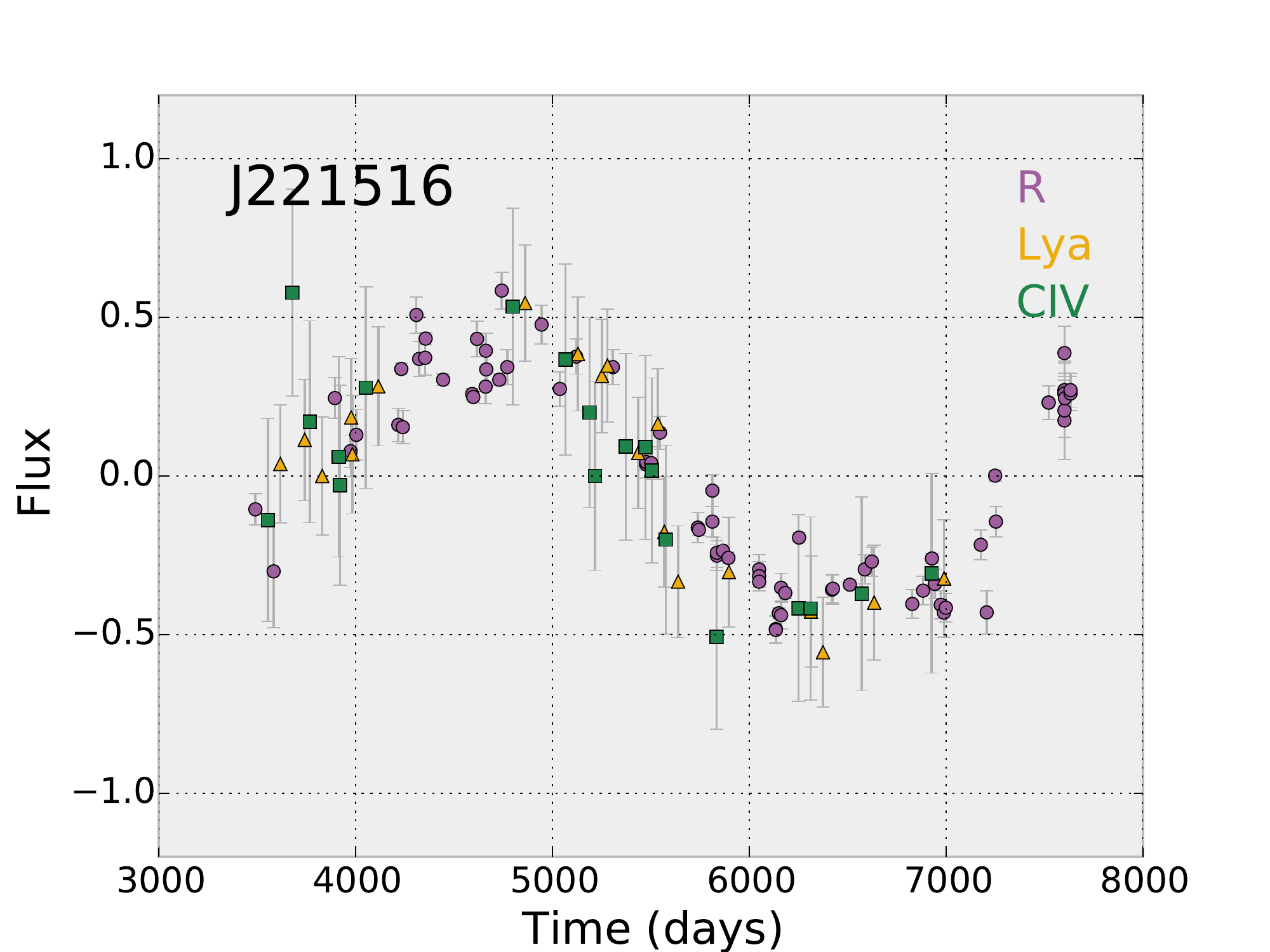}
\end{center}
\caption{Shifted and normalized continuum and emission line light
  curves for CT286 (left) and J221516 (right). The time axis is
  expressed in Julian Dates - 2450000 days. R-band continuum,
  Ly$\alpha$, and CIV light curves are presented using circles,
  triangles and squares, respectively.}
\end{figure}

\section{Radius--luminosity relations}

As well as providing extremely valuable information about the
innerworks of the BLR (see next section), reverberation mapping has
provided a huge scientific legacy with the determination of the so
called radius--luminosity relations. This tight correlations between
the distance at which one particular emission line is produced and the
continuum luminosity of the central engine allows for the
determination of BH masses by applying these calibrations to a
virialized BLR (i.e., $M_{\rm BH} \propto R v^2$, where $R$ comes from
the radius--luminosity relation and $v$ is measured from the width of
the broad emission lines).

So far, RM of the H$\beta$ line for nearby ($z < 0.3$) AGN has
produced a solid radius--luminosity relation for this line
\citep{1999ApJ...526..579W, 2000ApJ...533..631K, 2005ApJ...629...61K,
  2006ApJ...644..133B, 2009ApJ...697..160B, 2013ApJ...767..149B}.
Cross--calibration of the correlation to other lines has been a
significant enterprise which has allowed to determine BH masses of
high-$z$ quasars whose Balmer lines are redshifted into the infrared
realm. In particular, MgII has proven to be a safe line to be used as
BH mass estimator \citep{2004MNRAS.352.1390M,2012MNRAS.427.3081T},
while it has been extensively shown that CIV yields unreliable results
\citep{2005MNRAS.356.1029B, 2007ApJ...671.1256N, 2008ApJ...680..169S,
  2012ApJ...753..125S, 2016MNRAS.460..187M}.

Our monitoring effort has provided a sizeable number of Ly$\alpha$ and
CIV lags at the high-luminosity end of the quasar distribution. This,
together with other measurements found in the literature for lower
luminosity AGN allows us now to determine radius--luminosity relations
for these lines. These are presented in Figure 3, while the analytical
expressions are as follows:

\begin{equation}
\frac{R_{\rm Ly\alpha}}{10\ {\rm lt-days}} = (0.52\pm0.59) \left [ \frac{\lambda L_{\lambda} (1345{\rm \AA})}{10^{43}\ {\rm ergs/s}} \right ]^{(0.45\pm0.22)}
\end{equation}

\begin{equation}
\frac{R_{\rm CIV}}{10\ {\rm lt-days}} = (0.24\pm0.08) \left [ \frac{\lambda L_{\lambda} (1345{\rm \AA})}{10^{43}\ {\rm ergs/s}} \right ]^{(0.52\pm0.06)}
\end{equation}

where $R_{\rm line}$ is the measured lag and $\lambda L_{\lambda}
(1345{\rm \AA})$ corresponds to the product $\lambda \times
L_{\lambda}$ as measured at 1345 \AA\ from each spectra. As can be
seen from Equation 1, the zero point and slope of the Ly$\alpha$
radius--luminosity correlation are poorly constrained. The reason is
clear after an inspection of Figure 3 which reveals that Seyfert type
sources show a large dispersion around the best fitted correlation, in
contrast with the situation for CIV. In fact, the new CIV
radius--luminosity relation is very close to that reported by
\citep{2007ApJ...659..997K}, who found a zero point of $0.24 \pm 0.06$
and a slope of $0.55 \pm 0.04$. Notice that the Ly$\alpha$ and CIV
relations are consistent within the errors.

\begin{figure}[h!]
\begin{center}
\includegraphics[scale=0.4]{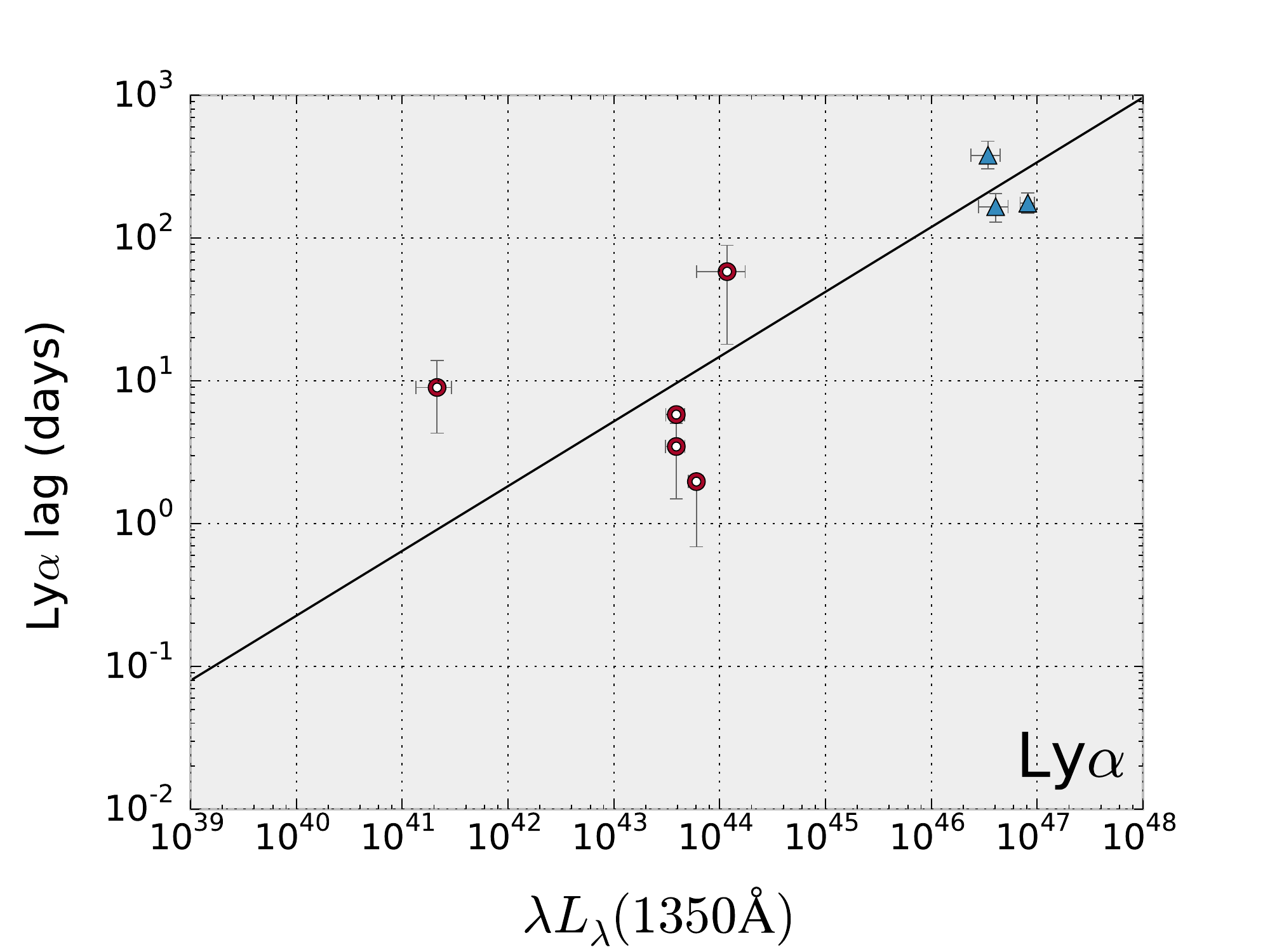}
\includegraphics[scale=0.4]{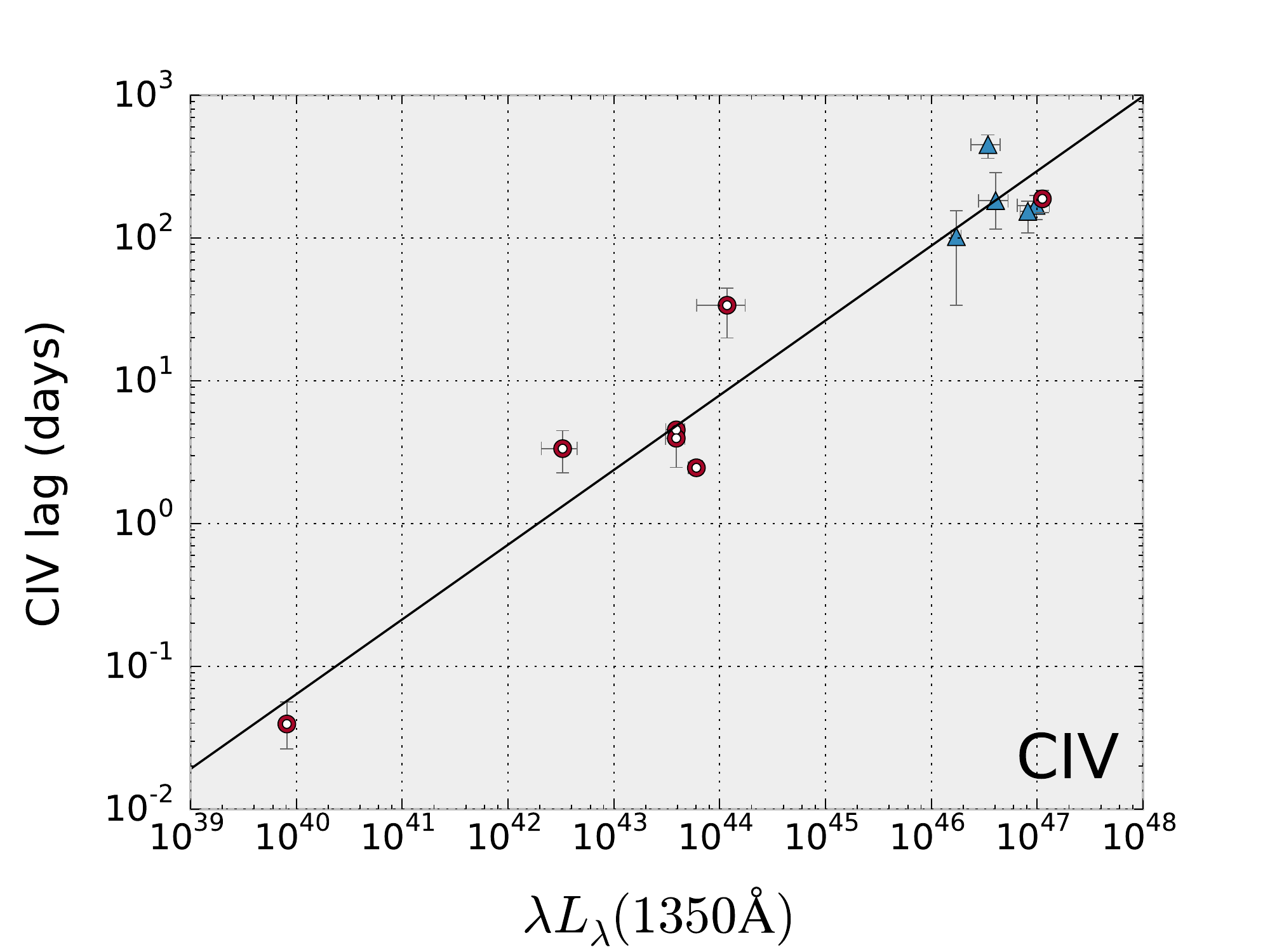}
\end{center}
\caption{Ly$\alpha$ (left) and CIV (right) radius--luminosity
  relations. Blue triangles correspond to results from our high-$z$,
  high-luminosity sample, while red circles are taken from the
  literature. These are NGC3783, NGC5548, NGC7469, F9 and 3C390.3 for
  the Ly$\alpha$ relation, and NGC4395, NGC5548, NGC3783, NGC7469, and
  3C390.3 for the CIV relation (\citeauthor{2005ApJ...632..799P},
  2005, \citeauthor{2006ApJ...641..638P}, 2006;
  \citeauthor{2015ApJ...806..128D}, 2015; Lira et al.,
  submitted). Taken from Lira et al.~(submitted), \copyright AAS.}
\end{figure}

\section{BLR stratification}

One crucial result that emerged early during the reverberation
campaigns of nearby AGN was that the BLR was compact, dense and
stratified \citep{1994ASPC...69....1P}, in contrast with previous
photoionization results that attempted to explain all emission lines
as produced by one set of physical parameters (or one single {\em
  cloud}). RM results made clear that different regions, with
different properties, and located at different distances from the
central engine, were producing the observed set of emission lines. To
build a consistent picture of the BLR, therefore, it is important to
determine where different lines are produced.

Our results allows to put constraints on the distance at which
Ly$\alpha$ and CIV are produced by obtaining $R_{\rm Ly\alpha}/R_{\rm
  CIV}$ for all source for which both lags have been measured. This is
presented in Figure 4, where three sources at high-luminosities come
from our lag determinations. Figure 4 clearly supports that $R_{\rm
  Ly\alpha}/R_{\rm CIV} \sim 1$ and that this ratio is independent of
luminosity.

\begin{figure}[h!]
\begin{center}
\includegraphics[scale=0.4]{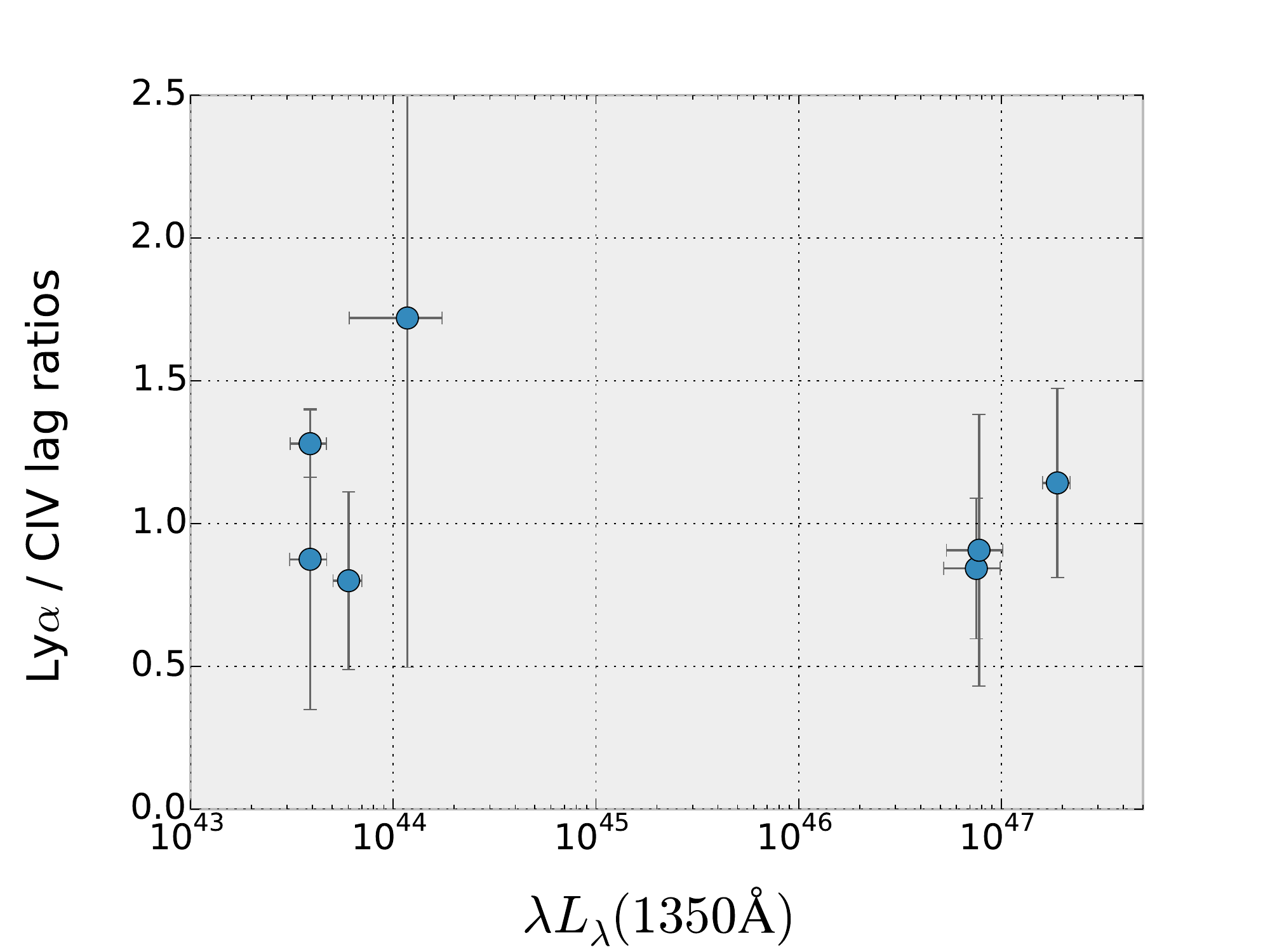}
\end{center}
\caption{Ly$\alpha$ to CIV lag ratios as a function of the $\lambda L_{\lambda}
(1345{\rm \AA})$ luminosity continuum.}
\end{figure}

\section{Summary and conclusions}

We have presented selected results from the RM campaign of 17 high-z,
high-luminosity quasars, which lasted more than 10 years. For several
sources lags between the continuum and BLR line emission were
determined, allowing us to extend radius-luminosity relationships up
to $\lambda L_{\lambda} (1345{\rm \AA}) \sim 10^{47}$
ergs/s. Continuum and line light curves for all sources can be found
in Lira et al.~(submitted).

\section*{Author Contributions}

PL, IB, SK and HN comply with the requirements for authorship of this article.

\section*{Acknowledgments}

PL greatly acknowledges the support of the Chilean National TAC
(CNTAC) which during more than 10 years allocated telescope time to
conduct our reverberation campaign and to the funding by Fondecyt along
all these years, and in particular to Project \#1161184.

\tiny
\bibliographystyle{frontiersinSCNS_ENG_HUMS}
\bibliography{plira_v2}

\end{document}